\documentclass[a4paper,10pt]{article}
\usepackage[top=2cm,bottom=2.2cm,left=2.5cm,right=2.5cm]{geometry}
\usepackage{amssymb}
\usepackage{amsmath}
\usepackage{epsfig}
\usepackage{hyperref}
\usepackage{apacite}
\usepackage{setspace}
\usepackage{authblk}
\usepackage{url}

\usepackage{array}
\usepackage{enumerate}
\usepackage{url}
\usepackage{verbatim}
\usepackage{ctable}

\newcommand{\parab}[1] {\bigskip \noindent \textbf{#1}~}
\newcommand{\param}[1] {\medskip \noindent \textbf{#1}~}

\newcommand{\mc} {\multicolumn}

\newfont{\helvetica}{phvr8t at 10pt}
\newfont{\helveticasmall}{phvr8t at 9pt}
\newfont{\helveticabig}{phvr8t at 22pt}
\newfont{\helveticao}{phvro8t at 11pt}

\ifdefined\labelenumiii
  \renewcommand{\labelenumiii}{(\roman{enumiii})}
\else
  \newcommand{\labelenumiii}{(\roman{enumiii})}
\fi

\usepackage{latexsym}

\title{Improving Query Expansion Using WordNet}
\author[1]{Dipasree Pal \thanks{dipasree.pal.gmail.com; Corresponding author. Fax.: +91 33 2577 3035; Tel.: +91 33 2575 2858. }}
\author[1]{Mandar Mitra \thanks{mandar.mitra@gmail.com; Fax.: +91 33 2577 3035; Tel.: +91 33 2575 2858.}}
\author[2]{Kalyankumar Datta \thanks{kalyankumardatta@gmail.com; Fax.: +91 33 2577 3035; Tel.: +91 33 2414 6129}}

\affil[1]{Indian Statistical Institute, 203 B.T. Road, Kolkata-700108, India}
\affil[2]{Jadavpur University, 188, Raja SCM Road, Kolkata-700032, India}
\date{}

\begin{document}
\maketitle


\begin{abstract}
  This study proposes a new way of using WordNet for Query Expansion (QE).
  We choose candidate expansion terms, as usual, from a set of
  pseudo relevant documents; however, the usefulness of these terms 
  is measured based on their definitions provided in a
  hand-crafted lexical resource like WordNet. Experiments with a number of
  standard TREC collections show that this method outperforms existing
  WordNet based methods. It also compares favorably with established QE
  methods such as KLD and RM3. Leveraging earlier work in which a
  combination of QE methods was found to outperform each individual method 
  (as well as other well-known QE methods), we next propose a combination-based QE
  method that takes into account three different aspects of a candidate
  expansion term's usefulness: (i) its distribution in the pseudo relevant
  documents and in the target corpus, (ii) its statistical association with
  query terms, and (iii) its semantic relation with the query, as
  determined by the overlap between the WordNet definitions of the term and
  query terms. This combination of diverse sources of information appears
  to work well on a number of test collections, viz., TREC123, TREC5,
  TREC678, TREC robust new and TREC910 collections, and yields significant
  improvements over competing methods on most of these collections.
\end{abstract}

\param{Keywords:}
 Query Expansion, Term Distribution, Term Association, Lexical Resources, Candidate Expansion Term, Pseudo Relevant Documents.


\section*{Introduction}   
Query Expansion (QE) is a widely used technique that attempts to increase
the likelihood of a match between the query and relevant documents by
adding semantically related terms (called \emph{expansion terms}) to a
user's query. The expanded query is supposed to retrieve more relevant
documents, thereby improving overall performance.

The source of the expansion terms is an important issue in query expansion.
Researchers have explored a variety of sources for collecting these terms.
Expansion terms may be taken from the whole target collection, or from a
few documents retrieved at top ranks in response to the original query.
Recently, researchers have explored the idea of collecting expansion
terms from the Web~\cite{mult-sources}, Wikipedia~\cite{wiki-sig07}, query logs~\cite{ql} of search engines, etc. Once a
set of candidate terms is determined, the aforesaid resources may also be
used to determine the importance of these terms. Finally, a few of the
candidate expansion terms (CETs) are selected for inclusion in the expanded
query.

Lexical resources like Ontology~\cite{Onto} / WordNet \linebreak (\url{http://WordNet.princeton.edu/}) are
also sometimes used as the source of expansion terms. Lexical resources
used for query expansion may be constructed either manually (e.g.\
WordNet), or automatically (usually based on co-occurrence information).
Since automatically constructed thesauri are usually based on corpus
statistics, they may contain linguistic flaws. In contrast, resources like
WordNet that are handcrafted by experienced lexicographers are expected to
contain less noise.

Thus, WordNet promises to be a good source of candidate expansion terms.
A number of previous studies have shown, however, that WordNet does not
necessarily work as expected. Voorhees~\cite{voor93,voor94} reported that
queries expanded using WordNet yield very little improvement, and sometimes
result in degraded performance, compared to the original, unexpanded
queries. Recently, Fang~\cite{wordnet-fang} showed that QE using WordNet
results in improved performance within the axiomatic framework. This method
demonstrates that an appropriate use of WordNet can indeed help in getting
useful results via query expansion. 

In this study, we first propose a new and effective way of using WordNet
for Query Expansion. The two features that distinguish our approach from
earlier work on WordNet-based QE are the following.
\begin{itemize}
\item As in many traditional QE methods, we select terms from top retrieved
  documents as candidates, instead of obtaining candidate expansion terms
  from WordNet itself.
\item The weight of a term depends not only on the semantic similarity
  between the term and all query terms (as determined by WordNet), but also
  takes into account the term's rareness, as well as the similarity score
  of the top-retrieved document(s) in which the candidate term was found.
\end{itemize}
This method outperforms the existing WordNet based methods.

The main contribution of this paper is a robust combination method which
determines the usefulness of candidate expansion terms by considering
multiple sources of information. In previous work on QE, several different
ways of estimating the usefulness of CETs have been proposed. We have
recently found that these different approaches may be combined to determine
the usefulness of a term more reliably~\cite{kldlca}. In this study, we
extend this idea by looking at:
\begin{itemize}
\item the distribution of a CET in the (pseudo) relevant documents as
  compared to its distribution in the complete corpus;
\item information about the association of a term with the query terms;
\item the semantic relation between a term and the query terms as
  determined by WordNet.
\end{itemize}
Based on the hypothesis that a combination of three totally different QE
approaches may perform better than each individual method, we propose a new
method which considers all of the three aforesaid features of a term. 


Our experiments confirm that the proposed approach leads to improved
performance. 
We tested our
methods on TREC123, TREC5, TREC678, TREC robust (new) and TREC910
collections. Compared to the baseline (original queries), the proposed
method consistently yields improvements that are significant for all
the above collections. It also compares favorably with other state of the
art QE methods, and improves upon the results reported recently by
Fang~\citeyear{wordnet-fang}.

The rest of the article is organized as follows. We discuss related work in
the next section. The following section describes the proposed methods. The
section `Experimental Setup' provides details about the IR system used for
our experiments, the collections used, evaluation metrics, etc.
Experimental results are presented and discussed in the subsequent section.
Finally, we conclude in the last section.

\section*{Related work}
\label{rel-work}
Early work on automatic query expansion (AQE) dates back to the 1960s.
Rocchio's relevance feedback method~\cite{rocchio71} is still used in its
original and modified forms for AQE. The availability of the TREC
collections, and the widespread success of AQE on these collections
stimulated further research in this area. \citeA{qe-survey} provide a recent and comprehensive survey of AQE
techniques. We focus here on some WordNet based AQE methods. We also
discuss some important AQE techniques that are either distribution- or
association-based.

\subsection*{Use of WordNet}
\label{sec:rw-WordNet}
WordNet has been used both for QE and to disambiguate the sense of query
words \linebreak \cite{lesk86,adlesk,ban-ped}. Our main focus here is on the use of
WordNet for QE. 
A number of issues need to be addressed when using WordNet as a source of
CETs.
\begin{itemize}
\item If a query word occurs in multiple synsets, which synset(s) should be
  selected?
\item Once some synsets have been selected, which words should be added to
  the query? Should only synonyms contained in these synsets be added? Or
  should hyponyms / hypernyms / meronyms / holonyms also be considered?
\end{itemize}
In early work, \citeA{voor93,voor94} explored these questions.
Given a search topic, she added a ``list of hand-selected WordNet
synsets containing nouns germane to the topic.'' Unfortunately, no
significant improvements were obtained even after manually selecting
``useful'' synsets and adding them to the query.

\citeA{zdli} used WordNet for sense disambiguation of query
terms, and then added synonyms of query words to expand the query. On the
CACM collection, their method yields an improvement of about 7\%
in P@10 over the original unexpanded queries. \citeA{sliu} also
disambiguate query terms and add synonyms, hyponyms, words from the term's
WordNet gloss, etc. They tested their method on TREC9, TREC10, and TREC12
robust queries and obtained very good results compared to other results
reported for these datasets.

More recently, Fang~\citeyear{wordnet-fang} reported positive results for
WordNet-based query expansion within an axiomatic retrieval framework. In
the method described by Fang, the set of candidate expansion terms consists
of all words from all the synsets in which query terms occur. A CET is
actually selected on the basis of the vocabulary overlap between its
glosses and the glosses of query terms.

In almost all the earlier studies mentioned above, CETs are taken from WordNet
itself. In contrast, we choose CETs from top ranked documents, and then use
WordNet to determine the semantic relatedness between query words and the
candidate terms. In particular, if a term's gloss shares many words with
the query words' glosses, we hypothesize (as in \cite{lesk86} and
\cite{ban-ped}, and later \cite{wordnet-fang}) that this term is related to
the query, and is therefore a good CET.

\subsection*{Other query expansion approaches}
\label{sec:rw-qe}
We now turn to two other QE methods that are used in this study:
association based methods and distribution based methods. Both methods use 
the set of pseudo-relevant documents (PRD) as a source of expansion term.

\param{Association-based QE techniques.} Early work on association-based
AQE includes ``concept-based'' QE~\cite{qiu-frei} and
\emph{phrasefinder}~\cite{phrasefinder}. Both methods make use of term
co-occurrence information extracted from a corpus. Local context analysis
(LCA)~\cite{lca, lgca} is another well-known method that also selects
expansion terms based on whether they have a high degree of co-occurrence
with all query terms. However, in LCA, co-occurrence information is
obtained from a set of top-ranked documents retrieved in response to the
original query, rather than the whole target corpus. Relevance-based
language models~\cite{rblm} constitute another, more recent, co-occurrence
based approach. This method is based on the Language Modeling framework.
The query and relevant documents are all assumed to be generated from an
underlying \emph{relevance model}. This model is estimated based on (only)
the pseudo relevant documents for a particular query. This approach was
subsequently refined by ~\citeA{rm3}. The refinement,
called RM3, incorporates the original query when estimating the relevance
model. According to the comparative studies by \citeA{comp-prf} and \citeA{roch-prox}, RM3 is the most effective and robust among a
number of state-of-the-art AQE methods. RM3 is frequently used as a
baseline against which several recent QE methods have been
compared~\cite{roch-prox,posrel,concept-weight,good-term,inter-passage}.

\param{Distribution-based QE techniques.} As early as 1978,
\citeA{doszkocs} proposed the interactive use of an associative  
dictionary that was constructed based on a comparative analysis of term
distributions. Also well known is Robertson's analysis of term selection
for query expansion~\cite{robertson90}. More recently, \citeA{carpineto-kld} proposed an effective QE method based on
information theoretic principles. This method uses the Kullback-Leibler
divergence (KLD) between the probability distributions of terms in the
relevant (or pseudo-relevant) documents and in the complete corpus.

\citeA{amati-thesis} proposes a new distribution based method which
uses Bose-Einstein statistics. This method also calculates the divergence
between the distribution of terms in the pseudo relevant document set and a
random distribution.

\param{Combination-based techniques.}
Efforts have also been made to combine AQE methods in various ways to
improve retrieval effectiveness. \citeA{carpineto-mult-rank} 
combined the scoring functions of a number of methods, all of them
distribution-based, to obtain improvements. A different approach is adopted
by \citeA{mult-sources}, who use multiple sources for query
formulation. In addition to the target corpus, information from `ClueWeb (\url{http://boston.lti.cs.cmu.edu/clueweb09/})
Heading Text', `ClueWeb Anchor Text' and Wikipedia (Available at: \url{http://download.wikimedia.org/enwiki/}) is used. Additionally,
Google N-grams (Available from Linguistic Data Consortium catalog), and a query log (Available as a part of Microsoft 2006 RFP dataset) from MSN are used to determine term
weights.

WordNet has also been used as a part of combination-based methods. The work
by \citeA{sliu} (mentioned in `related work' section )
combines information from WordNet with information about the correlation
(or co-occurence) of words with query terms. Similarly, \citeA{wordnet-fang} also tested a combination method that uses WordNet
as well as the mutual information of terms. The combination-based method
performed better than the individual methods.

Our earlier work on combination-based QE method~\cite{kldlca} is different
in that it combines a distribution-based method with an association-based
method (based on our belief that these two classes of methods offer
different advantages). Further, rather than simply combining scores, we
used one method to refine the set of terms selected by the
other (Of course, this can also, strictly speaking, be regarded as
  a combination where one component is very highly weighted.). This approach
is somewhat similar in spirit to a method proposed by \citeA{sel-good-term}, in which terms selected using standard pseudo
relevant feedback (PRF) are refined using a classifier that is trained to
differentiate between useful and harmful candidate expansion terms. Our earlier
work is most strongly related to that of \citeA{compcomb}, who also combine co-occurrence-based and
distribution-based methods. In their work, the combination is relatively
straightforward, however: one method is used for term selection and the
other for weighting. Word co-occurrence is measured using the Tanimoto
coefficient. Distributional differences are measured based on KLD or
Bose-Einstein statistics. The methods are tested on a relatively small
Spanish dataset. We used the well-known LCA (new) and RM3 methods (instead of
Tanimoto coefficient) to quantify term association. Also, instead of simply
using one method for term selection and the other for weighting, we combine
both methods for selection. Finally, we tested our method on a number of
large TREC datasets.

In the present study, we extend our earlier work by first proposing a new
and effective way of using WordNet for QE, and then combining this WordNet
based method with association based and distribution based AQE methods to
improve overall performance.


%

\section*{Proposed Methods}
\label{sec:prop}


\subsection*{Proposed WordNet Approach (P-WNET)}
\label{sec:mod-wnet}
During query expansion, the first important decision is the choice of the
source of candidate expansion terms. A set of sample relevant
documents for a given query would be a good source of CETs. In the absence
of true relevance judgments, the $D$ top-ranked documents retrieved in
response to the given query may be regarded as a set of  pseudo relevant
documents (PRDs).
Many well-known QE methods~\cite{carpineto-kld,lca,dfr} have shown that PRDs
are a good source of CETs. In contrast, WordNet-based QE typically starts
with the synonyms (and possibly holonyms, meronyms, etc.) of the query words
as CETs. Instead of confining ourselves to this set, we consider all terms
from PRDs as CETs.

Earlier work has shown that two terms tend to be strongly related if their
WordNet definitions share many common terms. Thus, if the definition(s) of
a term shares words with query word definitions, then the term may be
semantically related to the query, even though it may not be a direct
synonym of (or otherwise explicitly related via WordNet relations to) query
words.
For example, consider a candidate term `spondylitis' for TREC query 604
(\emph{Lyme disease, arthritis}). WordNet definitions for the CET spondylitis, query term 
arthritis and the query phrase `Lyme disease' share the terms `inflammation' and
`joint'. Thus, spondylitis appears to be strongly related to the query.
Similarly, the term `ill' (whose definition contains `disease') is
found to be related to `Lyme disease'.

The above idea has been used to quantify the relationship between a CET $t$
and a query word $q_i$ as follows. The definitions of $t$ and $q_i$ are
considered as two sets of words, and the overlap between these two sets is
taken as $\mathit{Rel}(t,q_i)$, the semantic similarity between $t$ and
$q_i$. The overlap may be measured using either the Jaccard coefficient
(Equation~\ref{jac}), as in \cite{wordnet-fang,ban-ped}, or the Dice
coefficient (Equation~\ref{dice}). Here, $c_x$ denotes the number of
documents in which term $x$ occurs, and $c_{x,y}$ is the number of
documents in which $x$ and $y$ cooccur.
\begin{equation}
  \label{jac}
  \mathit{Rel}_{t,q_i} = \frac{c_{t,q_i}}{c_t + c_{q_i} - c_{t,q_i}}
\end{equation}
\begin{equation}
  \label{dice}
  \mathit{Rel}_{t,q_i} = \frac{2 *c_{t,q_i}}{c_t + c_{q_i}}
\end{equation}
Our experimental results show that Dice
co-efficient performs somewhat better than Jaccard co-efficient.

\smallskip\noindent{\textbf{Phrases.}} Phrases usually have a specific
meaning that goes beyond the sum of the meaning of the constituent words.
We therefore give more priority to phrases in the query than single words
when finding definitions from WordNet. Any two consecutive words are
considered as a potential phrase, and such pairs are looked up in WordNet
first. If the phrase does not have an entry, then its constituent words are
looked up separately. Note that we consider phrases only at the time of
finding definitions from WordNet; we do not consider phrases anywhere else.
Specifically, documents and queries are indexed using single words only.
Thus, all CETs are single words. WordNet definitions are also considered to
be bags of single words.

After computing the relationship between a CET $t$ and the query words, we
consider the importance of $t$ in the collection as a whole, as given by
its \emph{idf}. We use Robertson's \emph{idf}~\cite{idf} formula as shown
in Equation~\ref{new-idf} ($N$ denotes the total number of documents in the
collection, and $N_t$ denotes the number of documents in which term $t$
occurs). If the \emph{idf} turns out to be negative, we use a very small
number ($0.0001$) instead.
\begin{equation}
  \label{new-idf}
  idf_t = \max {\Bigg(}0.0001,  \log_{10}\frac{N-N_t+0.5}{N_t+0.5} {\Bigg)}  \\
\end{equation}

Next, we factor in the importance of the pseudo relevant documents in
which $t$ occurs. This is intended to capture the intuition that terms
coming from relevant document are better than the terms coming from
non-relevant documents. In a pseudo relevance feedback setting, this
translates to the hypothesis that terms coming from top ranked documents
are likely to be more useful as CETs. Accordingly, we modify the score of
$t$ by the normalized similarity of documents in which $t$ occurs.
Equation ~\ref{small-s} shows how all the above factors are combined.
\begin{equation}
\label{small-s}
  s(t, q_i) =  \mathit{Rel}_{t, q_i} * \mathit{idf}_t * \sum_{d \in \it{PRD}}{\Bigg(}\frac{\mbox{Sim}(d,Q)}{\underset{d^\prime \in PRD}{\operatorname{{max}~}}\mbox{Sim}(d^\prime,Q)}{\Bigg)}  \\
\end{equation}
Here, $\mbox{Sim}(d,Q)$ denotes the similarity score of document $d$
with respect to the query $Q$.

The actual score of a CET $t$ is given by $S(t)$ (Equation~\ref{big-s}).
The summand in Equation \ref{big-s} is a slowly growing and bounded function of
$s(t,q_i)$. This function ensures that the weight of a CET lies within a
small range of values.
\begin{equation}
  \label{big-s}
  S(t) = \sum_{q_i \in \mathit{Q}} \frac{s(t, q_i)}{1+s(t, q_i)}  \\
\end{equation}
The $T$ CETs with the highest $S(t)$ scores are selected for inclusion in
the expanded query. The weights of terms in the final expanded query are
obtained by combining the normalized weights of expansion terms
(Equation~\ref{s-exp}) along with the normalized weights of original query
terms (Equation~\ref{s-orig}), as shown in Equation~\ref{f-score}.
\begin{equation}
  \label{s-exp}
  score_{exp}(t) = \frac{S(t)}{\max\limits_{t^\prime \in d \in {PRD}} S(t^\prime) }
\end{equation}
\begin{equation}
  \label{s-orig}
  score_{orig}(t) = \frac{1+\log(\it{tf(t,Q)})}{1 + \max\limits_{t^\prime \in Q} \log(\it{tf(t',Q)})}
\end{equation}
\begin{equation}
  \label{f-score}
  score(t) = score_{exp}(t) + \beta * score_{orig}(t)
\end{equation}
Since our focus is on short, title-only queries that typically contain
2--3 query words with each term occurring only once,
Equation~\ref{s-orig} assigns a weight of 1 to all query terms most of the
time. However, since the original query terms are supplied by the user,
 we regard them as being more important compared to the
automatically added expansion terms (as in ~\cite{ori-q}). We therefore
set $\beta$ to $2$.

\subsection*{Combination Method (KLWNET)}
\label{sec:klwnet}
As explained in first section, our goal is to improve QE by combining three
very different methods for estimating the usefulness of a CET: (i)~a
distribution based method, (ii)~an association based method; and
(iii)~WordNet. In earlier work~\cite{kldlca}, we have already explored four
possible combinations of two distribution based (\emph{KLD}, \emph{Bo1})
and two association based methods (\emph{LCA}, \emph{RM3}): 
\emph{KLDLCA}, \emph{KLDRM3}, \emph{Bo1LCA}, and \emph{Bo1RM3}.
We select one of these methods (\emph{KLDLCA}) and combine it with
\emph{P-WNET} (mentioned in the previous subsection).

Our combination method is very simple. We included in the expanded query
all the terms suggested by the contributing methods. We use
Equation~\ref{mixed-score} to determine the weights of expansion terms.
\begin{equation}
  \label{mixed-score}
  score(t) = \alpha * score_1(t) + (1-\alpha) * score_2(t)
\end{equation}
where $score_1$ and $score_2$ are the normalized scores of a term computed using
\emph{P-WNET} and \emph{KLDLCA}, respectively. In this formula, the
parameter $\alpha$ is used to control the relative importance of
\emph{P-WNET} and \emph{KLDLCA}. We experimented with values of $\alpha$
ranging from 0.1 to 0.9 in increments of 0.1 on our training corpus
(details in the next section), and found $\alpha = 0.3$ works
well. This is consistent with the fact that, in isolation, \emph{KLDLCA}
performs better than \emph{P-WNET} (Table 4). 
Thus, a small value of alpha should be preferred. 

\section*{Experimental Setup}
\label{sec:exp-set}
Table 1 lists the details of the test collections used in
our experiments. Since the TREC678 collection consists of a large number of
queries (150), and is more recent compared to the TREC123 collection, it is
used as a ``training set'' for the purpose of tuning the parameters of our
methods. As real-life queries are very short, we used only the title field
of all queries. Many of the queries thus contain only one term, and most of
the remainder are no longer than three words.

\begin{table}
\scriptsize
\begin{center}

\ \\
{\small
\begin{tabular}{ccc} \hline
  Query Id. & \# of Queries & Documents\\\hline
  TREC123 & 150 & TREC disks 1, 2\\
  51--200 & &\\\hline
  TREC5 & 50 & TREC disks 2, 4\\
  251--300 & &\\\hline
  TREC678 & 150 & TREC disks 4, 5 - CR\\
  301--450 &&\\\hline
  ROBnew & 100  & TREC disks 4, 5 - CR\\
  601--700 &&\\\hline
  TREC910 & 100  & WT10G\\
  451--550 &&\\\hline
\end{tabular}
}
\end{center}
\caption{Test collections }
\label{query-stats}
\end{table}

We used the TERRIER (\url{http://terrier.org/}) retrieval system for our
experiments. At the time of indexing, stopwords are removed and Porter's
stemmer is used as preprocessing. All documents and queries are indexed
using single terms, no phrases are used. The IFB2 variant of the Divergence
From Randomness (DFR) model~\cite{dfr} --- a relatively recent model that
performs well across test collections --- is used for term-weighting in all
our experiments as it performs better compared to the other variants available
within TERRIER. Parameters are set to the default values used in TERRIER. 

Results are evaluated using standard evaluation metrics (Mean Average
Precision (MAP), Geometric Mean Average Precision (GM\_MAP), precision at
top 10 ranks (P@10), and overall recall (number of relevant documents
retrieved)). Additionally, for each expansion method, we report the
percentage of queries for which the method resulted in an improvement in
MAP of more than 5\% over the baseline (no feedback). This number may be
viewed as indicative of how safe or robust an expansion method is. We use 
a two-tailed paired $t$-test with a confidence level of 95\% to check 
for statistically significant differences.


\section*{Experimental Results}
\label{sec:exp-res}
The objectives of our experiments are two-fold.
\begin{itemize}
\item (Experiment 1) To explore the effectiveness of the
  proposed WordNet based QE method (P-WNET), and to compare it with the
  baseline as well as existing WordNet based methods. We analyze the
  results in order to estimate the contribution of each factor in
  Equation~\ref{small-s} to the overall retrieval effectiveness. We also
  compare the results with those obtained using two state-of-the-art
  methods, and show that our proposed approach compares favorably with
  these methods.
\item (Experiment 2) To evaluate KLWNET, which combines
  distribution based information, association based information and
  information from WordNet. This method is also compared to
  state-of-the-art QE methods.
\end{itemize}

\subsection*{Experiment 1: Proposed WordNet Approach (P-WNET)}
\label{subsec:pwnet}
We use the following baselines for comparison in Tables 2 and 3. 
To the best of our knowledge, no comparisons between
WordNet-based methods and other state-of-the-art QE methods such as
KLD~\cite{carpineto-kld} and RM3~\cite{rm3} have been presented in earlier
work. However, reported figures for WordNet-based QE techniques are
generally much lower than the figures reported for these methods. We
include these two methods in our list of baselines.
\begin{itemize}
\item No feedback. The original, unexpanded queries are used for retrieval
  using the baseline method described in the section `Experimental Setup'.
\item FANG. The results reported by Fang in~\cite{wordnet-fang} are included
  in Table 3 under this label.
\item FN-PW. This method is very similar to the approach described by
  \citeA{wordnet-fang}, except for the basic setup, which is adopted
  from P-WNET. The two differences between FN-PW and FANG are ~(i) in
  FN-PW, CETs are obtained from the top ranked documents, while Fang
  selects CETs from the synsets of query words; ~(ii) FN-PW uses the DFR
  model for term weighting, while Fang's work is based on the axiomatic
  framework. The motivation here is to study the effect of these changes on
  Fang's approach.
\item No-WNet. Recall from Equation~\ref{small-s} that P-WNET combines information from WordNet
  with (i) information about the idf of a term, and (ii) the goodness of
  the document(s) from which the candidate term is obtained. No-WNet omits
  the WordNet-based factor in Equation~\ref{small-s} and makes use of only
  the last two sources of information mentioned above.
\item KLD. We choose KLD~\cite{carpineto-kld} as a state-of-the-art
  representative of distribution-based AQE methods. The parameter settings
  chosen for this method are as follows.
  \begin{itemize}
  \item $D$, the number of top-ranked documents that are assumed to be
    relevant, is set to $10$;
  \item $T$, the number of expansion terms is set to $40$.
  \end{itemize}
  These settings are in agreement with the observations of \citeA{carpineto-kld}. Our experiments also confirm that these values
  work well for KLD across collections.

  Note that the results presented here correspond to our implementation of
  KLD within TERRIER. While our implementation provides better results than
  TERRIER's native implementation of KLD, we were not able to exactly
  replicate the results reported in~\cite{carpineto-kld}. This is likely
  due to differences between the retrieval functions, indexing or query
  processing. For example, using full queries (title, desc and narr) on the
  TREC8 collection, and BM25 as the base term-weighting formula, we get MAP
  scores of 0.2992 for KLD (compared to a baseline of 0.2625). When using
  the IFB2 model, however, the baseline is higher (MAP = 0.2753), but KLD
  appears less effective (MAP = 0.2850).
\item RM3. This is also a relatively recent state-of-the-art AQE method. We
  use $D = 50$ documents (as suggested in~\cite{rblm}) and $T = 50$ terms.
  We set the Dirichlet smoothing parameter ($\mu$) to 2500 and the
  interpolation parameter to 0.5, based on the default settings for these
  parameters in Lemur (~\url{http://www.lemurproject.org/}). As before,
  we used the TREC678 collection to verify that these parameter values work
  well for us. In fact, for a number of datasets, our results for RM3 are
  superior to those reported in other recent papers (\cite{concept-weight},
  for example).
\end{itemize}

\parab{Parameter settings.} For our proposed method \emph{P-WNET}, and for
the training set mentioned in the previous section, a choice of $D=10$
documents and $T=60$ terms works well. We used these same parameters for
the No-WNet method, as the main motivation behind this method is to test
the effectiveness of the Wordnet-based component in P-WNET. We also tested
various parameter combinations for the FN-PW method on the TREC678
collection, and found that these same values are appropriate for this method
as well. Parameter settings for the other methods (KLD and RM3) are
discussed above.

\begin{table*}[t]
  \scriptsize
  \begin{center}
    \begin{tabular}{cccccccc} \hline
     Dataset & Measure & Baseline & FN-PW & no-wnet & KLD & RM3 & P-WNET \\ \hline

      TREC123   &            MAP  &           0.218 &           0.242 &           0.262 &           \textbf{0.274} &           0.249 &        0.273$^{B,p,n,r}$\\ 
                &                 &                 &          (10.8) &          (20.0) &          (25.4) &          (14.1) &       (24.9)\\ 
                &         GM\_MAP  &           0.097 &           0.111 &           0.096 &           0.101 &           0.109 &        \textbf{0.112}\\ 
                &                 &                 &          (14.4) &          (-0.8) &           (4.8) &          (13.1) &       (15.9)\\ 
                &           P@10  &           0.481 &           0.515 &           0.525 &           \textbf{0.537} &           0.511 &        0.526\\ 
                &                 &                 &           (7.2) &           (9.3) &          (11.8) &           (6.2) &        (9.4)\\ 
                &     \#rel\_ret  &          16536  &          17345  &          17901  &          18299  &          17702  &       \textbf{18377} \\ 
                &                 &                 &           (4.9) &           (8.3) &          (10.7) &           (7.1) &       (11.1)\\ 
                &   $>$ baseline on  &              0  &             60  &             58  &             62  &             64  &          \textbf{68} \\ 
         \hline
        TREC5   &            MAP  &           0.157 &           0.164 &           0.154 &           0.168 &           \textbf{0.170} &        \textbf{0.170}$^n$\\ 
                &                 &                 &           (4.3) &          (-2.3) &           (6.9) &           (8.2) &        (8.4)\\ 
                &         GM\_MAP  &           0.043 & \textbf{0.047} &           0.030 &           0.035 &           0.045 &        0.042\\ 
                &                 &                 &          (11.3) &         (-28.5) &         (-18.4) &           (7.1) &       (-0.2)\\ 
                &           P@10  &           0.286 &           0.326 &           0.284 &           0.268 &  \textbf{0.336} &        0.310\\ 
                &                 &                 &          (14.0) &          (-0.7) &          (-6.3) &          (17.5) &        (8.4)\\ 
                &     \#rel\_ret  &           1936  &           2075  &           1945  &           2184  &           2077  &        \textbf{2383} \\ 
                &                 &                 &           (7.2) &           (0.5) &          (12.8) &           (7.3) &       (23.1)\\ 
                &   $>$ baseline on  &              0  &             48  &             32  &             42  &             \textbf{50}  &          44 \\ 
         \hline
      TREC678   &            MAP  &           0.218 &           0.233 &           0.234 &           \textbf{0.257} &           0.230 &        0.255$^{B,p,n,r}$\\ 
                &                 &                 &            (6.6) &           (7.5) &          (18.0) &           (5.6) &       (17.0)\\ 
                &         GM\_MAP  &           0.100 &           0.119 &           0.099 &           0.101 &           0.106 &        \textbf{0.125}\\ 
                &                 &                 &          (19.1) &          (-0.5) &           (1.3) &           (5.9) &       (25.5)\\ 
                &           P@10  &           0.431 &           0.448 &           0.429 &           0.438 &           0.435 &        \textbf{0.451}\\ 
                &                 &                 &           (3.9) &          (-0.6) &           (1.6) &           (0.8) &        (4.6)\\ 
                &     \#rel\_ret  &           7287  &           7638  &           7770  &           \textbf{8556}  &           7617  &        8246 \\ 
                &                 &                 &           (4.8) &           (6.6) &          (17.4) &           (4.5) &       (13.2)\\ 
                &   $>$ baseline on  &              0  &             54  &             49  &             52  &             45  &          \textbf{55} \\ 
         \hline
       ROBnew   &            MAP  &           0.278 &           0.296 &           0.302 &           0.312 &           0.305 &        \textbf{0.321}$^{B,p,n}$\\ 
                &                 &                 &           (6.5) &           (8.6) &          (12.2) &           (9.8) &       (15.4)\\ 
                &         GM\_MAP  &           0.179 &           0.198 &           0.171 &           0.182 &           0.199 &        \textbf{0.200}\\ 
                &                 &                 &          (10.5) &          (-4.6) &           (2.0) &          (11.2) &       (11.7)\\ 
                &           P@10  &           0.421 &           0.432 &           0.437 &           0.405 &           \textbf{0.442} &        0.437\\ 
                &                 &                 &           (2.6) &           (3.8) &          (-3.8) &           (5.0) &        (3.8)\\ 
                &     \#rel\_ret  &           2887  &           3050  &           3082  &           \textbf{3172}  &           3002  &        3143 \\ 
                &                 &                 &           (5.6) &           (6.8) &           (9.9) &           (4.0) &        (8.9)\\ 
                &   $>$ baseline on  &              0  &             52  &             53  &             52  &             56  &          \textbf{58} \\ 
         \hline
        TREC910   &            MAP  &           0.195 &           0.206 &           0.188 &           0.193 &           0.211 &        \textbf{0.213}$^n$\\ 
                &                 &                 &           (5.5) &          (-4.0) &          (-1.1) &           (8.0) &        (9.0)\\ 
                &         GM\_MAP  &           0.081 &           \textbf{0.089} &           0.055 &           0.056 &           0.087 &        0.081\\ 
                &                 &                 &          (10.0) &         (-32.2) &         (-30.2) &           (7.4) &        (0.2)\\ 
                &           P@10  &           0.307 &           0.331 &           0.310 &           0.293 &           0.329 &        \textbf{0.336}\\ 
                &                 &                 &           (7.7) &           (1.0) &          (-4.6) &           (7.0) &        (9.3)\\ 
                &     \#rel\_ret  &           3770  &           3916  &           3440  &           \textbf{3987}  &           3889  &        3981 \\ 
                &                 &                 &           (3.9) &          (-8.8) &           (5.8) &           (3.2) &        (5.6)\\ 
                &   $>$ baseline on  &              0  &             \textbf{55}  &             38  &             44  &             53  &          47 \\ 
         \hline
    \end{tabular}
  \end{center} 
  \caption{Improvements obtained using P-WNET on different datasets. The ``$>$
    baseline on'' line shows the \textbf{\%-age} of queries for which each
    method beats the baseline by $>5\%$. Superscripts $B$, $p$, $n$, $k$,
    $r$ denote a statistically significant improvement of the proposed
    method over the baseline (no feedback), FN-PW, no-wnet, KLD and RM3
    respectively. For a particular collection, the highest value for any
    metric is shown in bold.}
\label{tab-wnet}
\end{table*}

\bigskip
From Table 2, it is clear that the proposed method P-WNET is
significantly better than the ``No feedback'' baseline on all collections.
Of course, our main concern is to compare our method with other
state-of-the-art QE methods. Our results show that P-WNET performs better
than FN-PW --- a method that is very similar to Fang's in terms of technique
as well as performance --- on all collections, with the improvements being
significant for the TREC123, TREC678, and ROBnew collections. P-WNET also
does better than RM3 on all collections, with the difference being
significant for the TREC123 and TREC678 collections. P-WNET is comparable
to KLD, with neither method being significantly better than the other.

\begin{table*}
  \scriptsize
  \begin{center}
    \begin{tabular}{ c  c  c  c  c  c  c  c }\hline
      Dataset & Baseline & Baseline     & FN-PW  & KLD    & RM3	   & P-WNET & FANG         \\ 
              &          & (Fang, 2008) &        &        &        &        & (Fang, 2008) \\ \hline
      TREC7   & 0.1891   & 0.1860       & 0.2020 & 0.2568 & 0.2025 & 0.2437 & 0.2160       \\
      TREC8   & 0.2467   & 0.2500       & 0.2622 & 0.2861 & 0.2591 & 0.2853 & 0.2660       \\ \hline
    \end{tabular}
  \end{center} 
  \caption{Comparison between Fang's WordNet-based method and P-WNET.}
  \label{comp-fang-wnet}
\end{table*}

Table 3 presents a direct comparison between P-WNET and
FANG on the two collections used in~\cite{wordnet-fang}. It shows that our
``no feedback'' baseline is close to the baseline used by Fang, but P-WNET
yields MAP values that are a good deal higher than the figures reported in
her paper. To make comparisons easy, MAP values for KLD and RM3 are also
included in Table 3. While FANG does slightly better
than RM3, KLD yields the best figures; these are noticeably higher than
those for FANG, but not significantly better than the figures for P-WNET.


\subsection*{Experiment 2: Combination Method (KLWNET)}
\label{subsec:klwnet}
We now turn to our combination based method. We compare KLWNET to KLD, RM3,
P-WNET and KLDLCA~\cite{kldlca}, a combination-based QE method that we
proposed in earlier work (see the `Related Work' section). In KLDLCA, candidate
expansion terms are first obtained (and weighted) using a distribution
based method (KLD); these terms are reranked (but not reweighted) based on
local context analysis; the top terms from the reranked list are included
in the final expanded query.

\param{Parameter settings.} For KLDLCA, we set $D=50$ documents and $T=40$
terms, since these values work well on our training dataset. For the KLWNET
method, which combines KLDLCA with P-WNET, we use the best parameter
settings for each of the constituent methods, i.e., $D=50$ documents and
$T=40$ terms for KLDLCA, and $D=10, T=60$ for P-WNET (see `Experiment 1' section).

\begin{table*}[t]
\scriptsize
  \begin{center}
    \begin{tabular}{cccccccc}
      \hline
      Dataset 	& 	Measure   & 	Baseline    & 		KLD   & 	RM3     & 	P-WNET      & 	KLDLCA      & KLWnet \\ \hline
      TREC123   &            MAP  &           0.218 &           0.274 &           0.249 &           0.273 &           0.283 &        \textbf{0.290}$^{B,k,r,w,kl}$\\ 
                &                 &                 &          (25.4) &          (14.1) &          (24.9) &          (29.3) &       (32.8)\\ 
                &        GM\_MAP  &           0.097 &           0.101 &           0.109 &           0.112 &           0.105 &        \textbf{0.114}\\ 
                &                 &                 &           (4.8) &          (13.1) &          (15.9) &           (8.4) &       (17.6)\\ 
                &           P@10  &           0.481 &           0.537 &           0.511 &           0.526 &  \textbf{0.567} &        0.550\\ 
                &                 &                 &          (11.8) &           (6.2) &           (9.4) &          (17.9) &       (14.4)\\ 
                &     \#rel\_ret  &          16536  &          18299  &          17702  &          18377  &          18850  &       \textbf{19138} \\ 
                &                 &                 &          (10.7) &           (7.1) &          (11.1) &          (14.0) &       (15.7)\\ 
                &$>$ baseline on  &              0  &             62  &             64  &             68  &             65  &          \textbf{69} \\ 
         \hline
        TREC5   &            MAP  &           0.157 &           0.168 &           0.170 &           0.170 &           0.171 &        \textbf{0.177}$^{B,k,kl}$\\ 
                &                 &                 &           (6.9) &           (8.2) &           (8.4) &           (9.0) &       (12.6)\\ 
                &        GM\_MAP  &           0.043 &           0.035 &           \textbf{0.045} &           0.042 &           0.036 &        0.040\\ 
                &                 &                 &         (-18.4) &           (7.1) &          (-0.2) &         (-14.6) &       (-5.6)\\ 
                &           P@10  &           0.286 &           0.268 &           \textbf{0.336} &           0.310 &           0.274 &        0.306\\ 
                &                 &                 &          (-6.3) &          (17.5) &           (8.4) &          (-4.2) &        (7.0)\\ 
                &     \#rel\_ret  &           1936  &           2184  &           2077  &           \textbf{2383}  &           2218  &        2294 \\ 
                &                 &                 &          (12.8) &           (7.3) &          (23.1) &          (14.6) &       (18.5)\\ 
                &$>$ baseline on  &              0  &             42  &             50  &             44  &             52  &          \textbf{56} \\ 
         \hline
      TREC678   &            MAP  &           0.218 &           0.257 &           0.230 &           0.255 &           0.266 &        \textbf{0.271}$^{B,k,r,w,kl}$\\ 
                &                 &                 &          (18.0) &           (5.6) &          (17.0) &          (22.0) &       (24.4)\\ 
                &        GM\_MAP  &           0.100 &           0.101 &           0.106 &           \textbf{0.125} &           0.103 &        0.117\\ 
                &                 &                 &           (1.3) &           (5.9) &          (25.5) &           (3.7) &       (17.2)\\ 
                &           P@10  &           0.431 &           0.438 &           0.435 &           \textbf{0.451} &           0.441 &        0.446\\ 
                &                 &                 &           (1.6) &           (0.8) &           (4.6) &           (2.2) &        (3.4)\\ 
                &     \#rel\_ret  &           7287  &           8556  &           7617  &           8246  &           8567  &        \textbf{8658} \\ 
                &                 &                 &          (17.4) &           (4.5) &          (13.2) &          (17.6) &       (18.8)\\ 
                &$>$ baseline on  &              0  &             52  &             45  &             55  &             57  &          \textbf{64} \\ 
         \hline
       ROBnew   &            MAP  &           0.278 &           0.312 &           0.305 &           0.321 &           0.326 &        \textbf{0.335}$^{B,k,r,w,kl}$\\ 
                &                 &                 &          (12.2) &           (9.8) &          (15.4) &          (17.2) &       (20.4)\\ 
                &        GM\_MAP  &           0.179 &           0.182 &           0.199 &           0.200 &           0.191 &        \textbf{0.203}\\ 
                &                 &                 &           (2.0) &          (11.2) &          (11.7) &           (6.8) &       (13.4)\\ 
                &           P@10  &           0.421 &           0.405 &           0.442 &           0.437 &           0.438 &        \textbf{0.453}\\ 
                &                 &                 &          (-3.8) &           (5.0) &           (3.8) &           (4.1) &        (7.4)\\ 
                &     \#rel\_ret  &           2887  &           3172  &           3002  &           3143  &           3173  &        \textbf{3194} \\ 
                &                 &                 &           (9.9) &           (4.0) &           (8.9) &           (9.9) &       (10.6)\\ 
                &$>$ baseline on  &              0  &             52  &             56  &             58  &             55  &          \textbf{60} \\ 
         \hline
        TREC910   &            MAP  &           0.195 &           0.193 &           0.211 &           0.213 &           0.204 &        \textbf{0.222}$^{B,k,kl}$\\ 
                &                 &                 &          (-1.1) &           (8.0) &           (9.0) &           (4.7) &       (13.5)\\ 
                &        GM\_MAP  &           0.081 &           0.056 &           \textbf{0.087} &           0.081 &           0.063 &        0.073\\ 
                &                 &                 &         (-30.2) &           (7.4) &           (0.2) &         (-22.4) &       (-9.2)\\ 
                &           P@10  &           0.307 &           0.293 &           0.329 &           \textbf{0.336} &           0.313 &        0.319\\ 
                &                 &                 &          (-4.6) &           (7.0) &           (9.3) &           (2.0) &        (4.0)\\ 
                &     \#rel\_ret  &           3770  &           3987  &           3889  &           3981  &           4021  &        \textbf{4159} \\ 
                &                 &                 &           (5.8) &           (3.2) &           (5.6) &           (6.7) &       (10.3)\\ 
                &$>$ baseline on  &              0  &             44  &             \textbf{53}  &             47  &             51  &          \textbf{53} \\ 
         \hline

 \end{tabular}
  \end{center}
  \caption{Improvements obtained using KLWNET on different datasets. The ``$>$
    baseline on'' line shows the \textbf{\%-age} of queries for which each
    method beats the baseline by $>5\%$. Superscripts $B$, $k$, $r$, $w$,
    $kl$ denote a statistically significant improvement of the proposed
    method over the baseline (no feedback), KLD, RM3, P-WNET and KLDLCA
    respectively. For a particular collection, the highest value for any
    metric is shown in bold.}
 \label{tab-klwnet}
\end{table*}

\medskip Table 4 shows that KLWNET yields the highest MAP
among all methods on all the datasets. On all collections, KLWNET also
emerges as the ``safest'' or most robust method, in the sense that it is
best in terms of the number of queries for which expansion improves
retrieval effectiveness. Additionally, KLWNET has the highest recall at
1000 documents for all corpora except TREC5, where KLWNET is second to
P-WNET. For the other two measures (GM\_MAP, P@10), KLWNET is consistently
among the top 3 methods, and differs very little from the best method.
Overall, KLWNET seems to perform very well on all measures and for all
collections.

%

\begin{table*}
  \scriptsize
  \begin{center}
    \begin{tabular}{ c  c  c  c  c  c  c } \hline
      Dataset & Baseline & Baseline     & KLD    & RM3    & KLWNET  & $\mathit{MII}_\mathit{mp}$  \\ 
              &          & (Fang, 2008) &        &        &         & (Fang, 2008)               \\\hline
      TREC7   & 0.1891   & 0.1860       & 0.2568 & 0.2025 & 0.2688  & 0.2370                     \\
      TREC8   & 0.2467   & 0.2500       & 0.2861 & 0.2591 & 0.3013  & 0.2800                      \\\hline
    \end{tabular}
  \end{center}
  \caption{Comparison between Fang's combination-based method and KLWNET.}
  \label{comp-fang-comb}
\end{table*}

Table 5 (similar to Table 3) presents
a comparison between KLWNET, RM3, KLD, and $\mathit{MII}_\mathit{mp}$, a
method proposed in \cite{wordnet-fang} that combines cooccurrence
information with information from WordNet. The figures for
$\mathit{MII}_\mathit{mp}$ are taken from \cite{wordnet-fang}. This method ($\mathit{MII}_\mathit{mp}$)
performs the best among all methods proposed in that paper. The table shows
that KLWNET's MAP figures are substantially higher than those for
$\mathit{MII}_\mathit{mp}$. KLWNET also outperforms KLD and RM3 on the
TREC7 and TREC8 collections.


\subsection*{Discussion}

The results in the preceding subsection confirm our hypothesis that, on
average, the combination based method works well. On all collections,
KLWNET outperforms both KLDLCA and P-WNET. A closer look at these results
(see Table 6) shows that, of the 150 queries in the TREC678 collection,
KLWNET does best on 40 queries. For the remaining queries, either KLDLCA or
P-WNET performs noticeably poorly. For each of these queries, the combination
manages to achieve an intermediate level of performance thanks to the
contribution of the superior method. Thus, the MAP obtained using KLWNET
lies between the figures corresponding to P-WNET and KLDLCA. Below, we
discuss some queries for which this pattern is observed.

\begin{table}
\scriptsize
  \centering
  \begin{tabular}[h]{l r}\hline
    Methods                      & \# queries \\\hline
    KLWNET $>$ KLDLCA, P-WNET    & 40         \\ 
    P-WNET $\ge$ KLWNET $\ge$ KLDLCA & 62     \\ 
    KLDLCA $\ge$ KLWNET $\ge$ P-WNET & 49     \\ \hline
  \end{tabular}
  \caption{Query-level comparison of various methods. (For one query, all
    methods perform equally well; this query is included in the counts in
    the last two rows of the table.)}
  \label{tab:disc}
\end{table}

\parab{P-WNET $\ge$ KLWNET $\ge$ KLDLCA.} For about half of these queries,
the original query terms appear to be vital. Since P-WNET attaches
relatively greater importance to the original query terms (see
Equation~\ref{f-score}), it does well on these queries. On the other hand,
when the query is expanded using KLDLCA, the importance of the original
query terms is diluted, resulting in a drop in performance (KLDLCA
  also makes use of Equations~\eqref{s-exp}--~\eqref{f-score} to compute
  term weights for the expanded query, but $\beta$ is set to 1, since this
  setting yields better overall performance.). Thus, for these queries, even
the baseline method yields higher MAP than KLDLCA. Since KLDLCA is the
dominant member of the combination method (see Equation~\ref{mixed-score}),
KLWNET also ends up performing poorly compared to P-WNET.

\begin{table}
  \scriptsize
  \centering
  \begin{tabular}{l c c c}\hline
                                 & P-WNET & KLDLCA & KLWNET \\\hline
    Wt.\ of original query term  &  2.94  &  1.00  &  1.00  \\
    \mc{1}{c}{(\emph{polygamy})} &        &        &        \\[2ex]
    Wt.\ of first exp.\ term     &  1.00  &  0.69  &  0.48  \\\hline
  \end{tabular}
  \caption{Weights of original and expansion terms for query 316}
  \label{tab:316}
\end{table}

Query 316 (Polygamy Polyandry Polygyny) is a typical example.
Table 7 compares the weights of original query terms with that
of the first (i.e. highest weighted) expansion term. When the query is
expanded using P-WNET, the original term \emph{polygamy} gets a weight of
2.94 (Equation~\ref{f-score}), while the first expansion term
(\emph{widow}) gets a weight of 1.00. Under KLDLCA, on the other hand, the
first expansion term (\emph{children}) gets a relatively high weight of
0.6865.
For KLWNET (Equation~\ref{mixed-score}), the first expansion term
(\emph{children}) gets a somewhat lower weight of 0.4806. Further, P-WNET
includes some some good terms like `widow', `monogami', and `polygamist' in
the expanded query with significant weights. Among these, only `widow' is
included in the expanded query by KLDLCA, but with a relatively low weight.


\parab{KLDLCA $\ge$ KLWNET $\ge$ P-WNET.} Query 361 (clothing sweatshops)
is an example query of the opposite kind. KLDLCA includes the useful term
`shop', but P-WNET does not. KLWNET therefore includes this term with a
lower weight. Both methods add the term `immigr' (a bad term), but KLDLCA
assigns a lower weight to this term compared to P-WNET. The combination
method assigns an intermediate weight to this term as well. Naturally,
KLWNET yields an Average Precision that lies in between the values
corresponding to KLDLCA and P-WNET. Unfortunately, however, no general
pattern seems to emerge for this class of queries.
 

%

\section*{Conclusion}
\label{sec:conclusion}
In this paper, we proposed a new way of using WordNet for Query Expansion.
This method outperforms the existing WordNet based methods. It also
compares favorably with established QE methods such as KLD and RM3. We
also proposed a combination of three QE methods that takes into account
different aspects of a candidate expansion term's usefulness. For each
candidate expansion term, this method considers its distribution, its
statistical association with query terms, and also its semantic relation
with the query. The combination of diverse sources of information appears
to work well, and yields results that are, on the whole, better than the
individual methods involved in the combination.

\bibliographystyle{apacite}
\bibliography{comb-wnet}


%
%
%

\end{document}